\documentclass[12pt]{article}
\usepackage{epsfig}
\date{~}

\begin{document}

\title{%
\vskip-6pt \hfill {\rm\normalsize UCLA/03/TEP/30} \\
\vskip-12pt~\\
Detectability of the Sgr Dwarf Leading Tidal Stream 
with Auger, EUSO or OWL}

\author{Graciela Gelmini\rlap{,}{$^{1}$} Paolo Gondolo\rlap{,}{$^{2}$}
  Adrian Soldatenko\rlap{$^{1}$}
  \\~\\
  \small \it ${}^{1}$Dept.\ of Physics and Astronomy, UCLA,
  Los Angeles, CA 90095\@.
  \\
   \small \it   Email: {\tt gelmini@physics.ucla.edu, asold@ucla.edu}\@.
  \\
  \small \it ${}^{2}$Dept.\ of Physics, University of Utah, 115 S 1400 E \#201,
  \\
  \small \it Salt Lake City, UT 84112-0830, USA. Email:
  {\tt paolo@physics.utah.edu}\@.
  \vspace{-2\baselineskip}
  }

\maketitle

\begin{abstract}
We point out that if heavy metastable particles composing the dark
matter of our galaxy are responsible for the ultra-high energy cosmic rays
 (UHECR) then the leading tidal stream of the Sagittarius dwarf galaxy
could be detected through UHECR.  The signal would be an anisotropy in
the UHECR flux smaller than the telltale anisotropy towards the
galactic center that would first establish unstable dark matter as the
origin of the UHECR.
\end{abstract}


Cosmic ray particles with energies above the Greisen-Zatsepin-Kuzmin 
cutoff \cite{GZK} of about $5\times 10^{19}$ eV have been detected by
a number of independent experiments over the last decade \cite{exp}.
The existence of these ultra-high energy cosmic rays (UHECR) presents
us with a problem.  Nucleons and photons with those energies have short
attenuation lengths and could only come from distances of 100 Mpc or
less, while plausible astrophysical sources for those energetic particles
are much farther away.  Recently, results from the HiRes 
Collaboration \cite{HiRes}
brought the violation of the GZK cutoff into question and this issue 
will only be resolved conclusively by the Pierre Auger Observatory.

Among the solutions proposed for the origin of the UHECR is the decay of
supermassive relic particles which may constitute all or a fraction of the
dark matter \cite{decayDM}.  Such particles must be metastable, with
lifetimes exceeding the age of the Universe.  In this scenario the
 flux $\phi$ of
UHECR coming from a particular volume element in the Universe is 
proportional to the density of dark matter in this volume, and inversely
proportional to the square of the distance $r$ between us and this particular
volume element. Thus the flux is the integral of the
column density over the solid angle 
\begin{equation}
\phi \sim \int \frac{d^3 r \rho(\vec r)}{r^2} = \int d\Omega \int
\rho(\vec r)dr .
\end{equation}
Consequently, in decaying dark matter models the UHECR are
 produced mostly in our
galaxy and the key test is the expected anisotropy in the arrival
directions caused by the offset of the Sun from the center of the galaxy
\cite{anisotropy}.  In these models cosmic rays from dark matter dominate at
energies above about 6$\times 10^{19}$~eV, and are mostly
 photons~\cite{aloisio}.
Even for protons the effects of magnetic fields on UHECR trajectories can
 be neglected at these energies~\cite{Stanev}, thus the
 trajectories of these
UHECR are straight lines and point towards the place of origin.
Spherical halo models yield an anisotropy symmetric around
 the direction of the Galactic Center (GC) (while triaxial models, which
 will not be considered here for simplicity, give considerable angular
 deviations).
The amplitude of this anisotropy is controlled by the halo core radius
$R_c$, a parameter which appears in the halo density models, the simplest of
which is the isothermal halo model
\begin{equation}
\label{eq:rhohalo}
\rho_{\rm halo} = \rho_o  \frac{R_c^2 }{r'^2+R_c^2}
\end{equation}
where $r'$ is the radius from the GC.

The halo core radius $R_c$ is not well determined from astrophysical means and
could in fact be measured through the anisotropy  itself. 
In isothermal halo models, the GC asymmetry amplitude, $A_{GC}(\theta)$,
defined as the ratio of the flux in the  direction of the GC and the 
flux from  the galactic anti-center (AC) within an integration cone
 of aperture $\theta$,
depends strongly on $R_c$. For example in
 \cite{Berezinsky}, for
$R_c =  5~$kpc, the amplitudes are found to be (see Fig.~5 of
 \cite{Berezinsky}) 
 $A_{GC}(5^\circ) \simeq 5$  and 
 $A_{GC}(80^\circ) \simeq 3 $, while for $R_c = 10$ kpc,
$A_{GC}(5^\circ) \simeq 3$ and $A_{GC}(80^\circ) \simeq 2$. 

 The observation of a  GC-AC asymmetry would prove the decaying 
particle scenario for the origin of the UHECR.
In this case, the leading tail of the Sagittarius Dwarf
 Galaxy would give rise to a 
different asymmetry in the UHECR fluxes. This is the new 
asymmetry we are considering here.

The Two Micron All
Sky Survey (2MASS) and the Sloan Digital Sky Survey (SDSS) 
 \cite{majewski,newberg} have traced the
tidal stream of the Sagittarius (Sgr) dwarf galaxy more than
$360^\circ$ around the sky.
This galaxy is a dwarf spheroidal  of roughly $10^9
M_\odot$. It is  a satellite of the  Milky Way Galaxy, 
located inside the Milky Way, $\sim$12 kpc behind the
Galactic Center (GC) and $\sim$12 kpc below the Galactic Plane (GP)
\cite{ibata97}.  There are two streams of matter that extend
outwards from the main body of the Sgr galaxy and wrap around the GC.
  These streams, known
as the leading and trailing tidal tails, are made of matter tidally
pulled away from the Sgr galaxy. They are on the orbital plane of the 
 Sgr galaxy, which crosses the Galactic plane near the position of 
the Solar system.
The leading tidal tail results from stars (and dark matter particles
[see \cite{paolo}]) that
were originally between the Sgr dwarf center and the center of the
Milky Way.  These stars that are closer to the GC
 orbit faster than the Sgr dwarf center, 
they move at lower gravitational
potential and have shorter orbital periods around the GC.  Thus, these
faster orbiting stars (and dark matter particles) stream out ahead 
of the main body of the Sgr dwarf
and create the leading tail.    The stars that give
rise to the leading tail are stripped at roughly a distance of 10 kpc
from the GC, on the opposite side of the GC from the Sun.  This
distance of 10 kpc is determined \cite{paolo} 
by subtracting the current distance
between the GC and the center of the Sgr main body, $\sim 16$ kpc,
minus the Sgr tidal radius, $\sim 6$ kpc.  This distance of 10 kpc
determines the perigalacticon of the leading tidal tail, both
where it was removed from the Sgr main body and again on this side
(the solar side) of the GC.  Similarly, the trailing
tail is formed from stars that were on the opposite side of the dwarf
when they were stripped, they are further out and thus move slower
than the Sgr galaxy main body.

 The trailing stream  roughly extends in a doughnut
 shape in the Southern
galactic hemisphere around the position of the Sun, at a distance of about 
20 kpc from us. 
The leading tail goes around the GC
in the Northern galactic hemisphere,  turns around at a radius of about
35 kpc, then approaches the galactic plane from the Northern galactic
 hemisphere and crosses this plane passing through the Solar system
(see Fig. 11 of \cite{majewski})
 in the general direction orthogonal to the Galactic plane. The streams
 carry dark matter as well as visible stars. Thus the leading stream brings
additional dark matter into the vicinity of the Solar system. The effect of
the leading  Sgr stream for direct detection experiments, if the DM
consists of weakly interacting massive particles, has recently been studied
in \cite{paolo} where the estimated range of DM density in the stream was
found to be
\begin{equation}
\rho_{\rm stream} = [2, 200] \times 10^4
M_\odot/{\rm kpc}^3 = [0.001, 0.08] \, {\rm GeV/cm^3}~.
\end{equation}
This density corresponds to a fraction $\chi$=(0.3-25)\% of the local
density of the isothermal Galactic halo for the usual estimate
 $\rho_{\rm h} = 0.3$ GeV/cm$^3$,
and to a fraction $\chi$=(0.45-37.5)\% of the local halo density
 for a lower estimate $\rho_{\rm h} = 0.2$ GeV/cm$^3$ (e.g. see the 
compilation of possible local halo density values for spherically
 symmetric halo models  in \cite{DAMA}).
This evaluation assumes a constant density along the stream \cite{paolo}.

 The trailing stream may lead to some
small increase of UHECR events coming from the Sgr dwarf plane,
 but this increase 
will be very small and would be more difficult to detect than
 the leading trail. Recall
that the flux depends on the integral of the density over the
 line of sight (see Eq.~(1)),
and we can see along the axis of the leading trail but only transversally
 through the trailing tail. 

 We  concentrate here on the leading tidal stream, which 
 passes through the solar system.  The width of
the tidal stream in 2MASS M stars is estimated to be 4-8 kpc and  
 the best-fit projection brings the center of the
leading tail within 2 kpc of the Sun \cite{majewski}.
 We can see in  Fig. 11 of \cite{majewski}
that the stream starts turning around at about (25-35)~kpc from the Sun,
 but for
smaller distances can be approximated by a cylinder.
 For simplicity  here we model the
leading stream  as a cylinder of radius $R$=(2-4)~kpc  and 
height $L$=(25-35)~kpc ``above" the plane of the galaxy, i.e. \
 on the side of
the North Galactic Pole (NGP). We assume that the central axis of 
the cylinder
passes through the position of the Sun. 
 The NGP is located at $\delta=+27^\circ~24'$ of declination and
$\alpha=$12$^h$ 49$^m$ of right ascension.

 We then compare  the expected UHECR flux
$\phi_\parallel$ in a solid angle of aperture $\theta$
 in the direction of the NGP (i.e. along the
axis of the cylinder), with the flux $\phi_\perp$ coming from an
identical cone taken in a direction perpendicular to both the directions
to the NGP and the GC (a cone with axis in the galactic plane). There are
two directions with this property, one pointing to
 $\delta = -48^\circ$, $\alpha=$9$^h$  and the other
pointing to $\delta=+48^\circ$, $\alpha=$21$^h$.  We define
the asymmetry amplitude
\begin{equation}
A_{\rm stream}(\theta) = \frac{\phi_\parallel}{\phi_\perp}.
\end{equation}

In a spherically symmetric halo without the Sgr stream 
there should be no difference 
in the fluxes from both directions, thus $A_{\rm stream}=$1. The presence
of the stream  leads to an asymmetry $A_{\rm stream} > $1.

Taking the evaluation of the density of the leading stream from
\cite{paolo} to be $\rho_{\rm stream} = \chi \rho_{\rm h}$, where
$\rho_{\rm h}$ is the halo density at the location of the Sun, and
assuming that the density is constant along the stream, we have
computed $A_{\rm stream}(\theta)$, using the halo model in Eq.~(2).
Notice that $A_{\rm stream}(\theta)$ depends on the local overdensity
$\chi$ due to the stream and not on the local halo density itself. In
Fig.~1 we plot the column density as seen from the location of the Sun
for the halo model in Eq.~(2) with $R_c$= 5 kpc and for several of our
leading stream models.  The increase in column density in the
directions close to the axis of the stream is clearly visible, except
for the lowest values of $\chi$.

In Figs.~2 and 3 we plot the asymmetry $A_{\rm stream}(\theta)$ for different
 values of $\chi$ and two values of $R_c=5$~kpc and 10~kpc 
(Fig.~2 and~3, respectively). 
With $R$ and $L$ as the dimensions chosen for the cylinder
representing the stream, $A_{\rm stream}(\theta)$ is constant for 
$\theta < \arctan(R / L)$, and decreases rapidly with increasing values of
$\theta$, as can be seen in Figs. 2 and 3.
 The maximum asymmetry,  $A_{\rm stream}(\theta \leq 6.5^\circ) = 1.7$,
is obtained for 
the largest cylinder, lowest halo density (i.e.\ highest $\chi$) 
and small core radius, namely for
 $R$=4~kpc, $L$=35~kpc, $\chi=$ 0.375 and $R_c$=5~kpc.
A larger core radius decreases the asymmetry to 
$A_{\rm stream}(\theta \leq 6.5^\circ) = 1.57$.
Reducing $L$ to 25~kpc one obtains 
$A_{\rm stream}(\theta \leq 9^\circ) = 1.5$ for $R_c=5$~kpc (and 1.38
 for $R_c=10$~kpc). 
Reducing the radius of the cylinder to $R$ = 3~kpc and reducing $\chi$
 to $\chi=0.25$, 
 with $L$=25~kpc the asymmetry amplitude becomes 
$A_{\rm stream}(\theta \leq 7^\circ) = 1.36$~(1.28)  for $R_c=$~5~(10)~kpc.
 These are some of the largest expected asymmetries, while
 the asymmetries  for the smallest values of $\chi$ are
negligible. For example, $A_{\rm stream} <1.005$ for
$R$=3~kpc, $L$=25~kpc and $\chi= 0.003$.

In order to detect the asymmetry $A_{\rm stream}(\theta)$
 above the three-sigma level we need
a minimum number of events $N$ within the two cones of aperture
 $\theta$ that we are comparing,
\begin{equation}
\frac{N_\parallel-N_\perp}{\sqrt{N_\parallel + N_\perp}}=
\sqrt{N} \frac{(A_{\rm stream} - 1) }{\sqrt{A_{\rm stream} + 1}} \, \geq \,  3,
\end{equation}
which means
\begin{equation}
N \, \geq \, 9 \, \frac{ (A_{\rm stream} + 1) }{ (A_{\rm stream} - 1)^2}.
\end{equation}

For the maximum value of the asymmetry, $A_{\rm stream}= 1.7$, we need at least
$N$=50 events; for $A_{\rm stream}= 1.4$, we need  $N$= 135 events; 
for $A_{\rm stream}= 1.2$, we need $N$= 500 events;
 for $A_{\rm stream}= 1.1$, we 
 need at least  $N$= few 10$^3$ events; and for  $A_{\rm stream}< 1.01$,
 we would need 
$N >$ a few  10$^5$ events. These are the number of events within the cones
 considered. While the number of events increases as the solid angle of the
 cones increases, the stream asymmetry decreases for cones of aperture
larger than $\arctan(R / L)$. Thus there is an optimum cone aperture
which maximizes the reach of each observatory.

Let us see how large an overdensity $\chi$ the Pierre Auger Observatory,
 EUSO and OWL would be able to measure. The reach of each observatory
is shown in Fig.~2 and Fig.~3 with thicker solid (red) lines.  
As can be seen in the figures,
 the integration
cones for which the  stream is best detectable through the asymmetry 
have an aperture $\theta$ of about $15^\circ$ to $20^\circ$.  For larger
 cones, although the minimum asymmetries that can be detected decrease with 
increasing number of events $N$ within each cone,
the predicted stream asymmetry decreases faster, so that the 
reach in $\chi$ does not improve.

To estimate the number of expected events, we need a value for the
UHECR flux.
At energies of $6 \times 10^{19}$~eV the AGASA and HiRes cosmic ray
 fluxes differ by about a factor of 3 (see e.g.\ Fig.~1 in
 \cite{owl}). Since it is the AGASA data that do not show a GZK
 cutoff, we adopt the AGASA value for the flux, with
 an energy dependence of $E^{-2}$, which is typical of decaying dark matter
models~\cite{aloisio}. This gives us an isotropic UHECR 
flux integrated above $6\times 10^{19}$~eV
of about 0.07~(km$^2$ sr yr)$^{-1}$. At higher energies the data points of
 AGASA, which are about one order of magnitude higher than those of HiRes,
 give an integrated isotropic flux  of about 0.04~(km$^2$ sr yr)$^{-1}$
for $E > 1\times 10^{20}$~eV.

The southern station of the Pierre Auger Observatory (see for example
 \cite{auger}) is already under 
construction since the year 2000 in the Province of Mendoza,
 Argentina, at a latitude of about 
 $35.2^\circ$ South.
 It will be completed by 2004. 
The northern station is proposed for Utah
and would then enable continuous full-sky coverage. 
 The southern array consists of 1600 particle detectors spread 
over 3000 km$^2$ with fluorescence
telescopes placed on the boundaries of the surface array. The expected angular 
resolution is less than 1$^\circ$ for all energies 
(0.5$^\circ$ at 10$^{20}$eV). This angular resolution is
very good  to localize the flux from the stream. The limiting effective
aperture for the full southern array  is 7350~km$^2$sr, including only
 zenith angles less than 60$^\circ$, which is the region of the sky where the 
reconstruction algorithm
 works best. Given that the latitude of the southern array is 
$35.2^\circ$ South, this means that only objects at declination
 smaller than $+24.8^\circ$ can be reconstructed well.
 However, the stream is centered at the NGP, which has
a declination of $+27.4^\circ$, and so it is outside the best detection
 region of Auger South.

Thus, the stream asymmetry
could not be studied by the Southern Auger site 
unless events below a
60$^\circ$ zenith angle can be analyzed effectively. In this case
the aperture would increase by about 50\%~\cite{auger}, 
becoming 11250~km$^2$sr, and
there would be a total of 790 events per year 
with $E \geq  6\times 10^{19}$~eV.
In the extreme case in which zenith angles as large as 85$^\circ$ 
could be reached, and we are not sure that this can be done, the number of 
events within the cones of solid angle $\Omega$ we are 
studying would be a fraction $f= \Omega/ (3.64\pi)= 
(1- \cos\theta)/(1+\cos 40^\circ)$ of the total number of events observed,
 and only cones with aperture 
$\theta < 22.4^\circ$ would be completely included in the
 observable portion of the sky. In this case
 the minimum $A_{\rm stream}(\theta)$
observable by Auger South in 10 years as function of $\theta$ for
$\theta < 22.4^\circ$
can be seen in Figs.~2 and 3 (thicker solid (red) line
 with the  label Auger S).
 In particular,
for  $\theta= 20^\circ$, we have $f= 0.034$, and after 10 years
 we would get $N= 268$ events with
 $E \geq 6 \times 10^{19}$~eV. This number of events would allow us to detect
 an asymmetry $A_{ \rm stream}(20^\circ) \geq$~1.28. From Figs.~4 and 5,
 which plot the overdensity $\chi$ due to the
 stream as a function of the asymmetry within a cone of $20^\circ$ aperture,
we see that if Auger South could observe at zenith angles of up to 85$^\circ$, 
it could detect the stream if $\chi$ is at least
$\chi =  0.3$ (this value occurs for $R$ = 4 kpc, 
$L$ = 35 kpc and $R_c = 5$ kpc).

For Figs.~4 and 5 we have chosen a cone aperture of $20^\circ$ because
this
aperture is close to that one that maximizes the reach, although the latter
depends somewhat on the observatory (see Figs.~2 and 3).

 If the Northern Auger site would become a reality,
the NGP and thus the stream would be in full view of this new
observatory. The projected total effective area of 
44000~km$^2$sr~\cite{arisaka}
of the  combined Southern and Northern Auger sites,
with total 4$\pi$~sr sky coverage, would yield 3080 events  with
 $E \geq 6 \times 10^{19}$~eV.
The number of events within one of the cones of solid angle $\Omega$ we are 
studying is a fraction $f= \Omega/ (4\pi)= 
(1-\cos\theta)/2$. For  $\theta= 20^\circ$,
 this gives $f=0.030$. Thus in 10 years
of combined Southern and Northern Auger data we would obtain
$N= 924$ events with
 $E \geq 6 \times 10^{19}$~eV. This would allow the detection of asymmetries 
above the thicker solid (red)
 line labeled Auger S$+$N shown in Figs.~2 and 3. In particular
 we could detect $A_{ \rm stream}(20^\circ) \geq$~1.14, which would translate
 into the detection of a stream
 with $\chi \geq  0.14$ (see Figs.~4 and 5).

The Extreme Universe Space Observatory (EUSO)~\cite{euso}, a 
 mission on-board the International
Space Station, will image the air fluorescence of the
 extensive air showers produced by
UHECR in the atmosphere. Its angular resolution is expected to be less than
2$^\circ$ and a conservative estimate of the duty cycle is 10\%.
 It will operate for three years, with an effective
trigger aperture which increases very fast with energy from
$1.0 \times 10^4$~km$^2$~sr  at $E = 4 \times 10^{19}$~eV to 
$4.5 \times 10^4$~km$^2$~sr at
 $E \geq 1 \times 10^{20}$~eV (these effective apertures include
the 10\% duty cycle).   The sky
 coverage is expected to be very close to complete, thanks to a suitable
 inclination of the EUSO detector axis with respect to the vertical and 
to the characteristics of the International Space Station orbit (the ISS
 orbit has a typical inclination of about $52^\circ$ with respect to the
 ecliptic, a period of about a 16th of a day, and a precession of the 
line of nodes of about $5^\circ$ per day~\cite{ISS}).
Thus, within $4\pi$~sr of sky coverage,
 we expect about 1800 events per year with 
$E \geq 1 \times 10^{20}$~eV.
The number of events within the cones of solid angle $\Omega$ we are 
studying is a fraction $f= \Omega/ (4\pi)=
(1- \cos\theta)/2$ ($f=0.030$ for  $\theta= 20^\circ$). This  means that
in three years
EUSO could detect asymmetries 
above the thicker solid (red) line labeled EUSO shown in Figs.~2 and 3.
In particular, in three  years of operation $N= 162$ events would be
 collected with $E \geq 1 \times 10^{20}$~eV 
for  $\theta= 20^\circ$. Thus, EUSO could
 measure a minimum asymmetry  $A_{ \rm stream}(20^\circ)=$~1.36, which is
 larger than the largest we obtained in our models at $\theta =20^\circ$
(see Figs.~2 and 3).

The Orbiting Wide-angle Light-collectors (OWL)~\cite{owl}
mission would consist of a
 pair of satellites observing extended air showers in the atmosphere
 produced by UHECR with stereoscopic view. The angular resolution would be
  less than 1$^\circ$. It would have an instantaneous
 aperture rapidly increasing with energy, from $4 \times 10^4$ km$^2$~sr at 
 $E = 3 \times 10^{19}$~eV to
 about $2 \times 10^6$ km$^2$~sr at $E \geq 1 \times 10^{20}$~eV.
 Taking into account that OWL
 can only view the dark side of the Earth, and accounting for the effect
 of the Moon, clouds and man-made light, the duty cycle is 
about 10\%. In fact the continuous effective aperture 
would be  $2.3 \times 10^5$ km$^2$~sr for $E \geq 1 \times 10^{20}$~eV~\cite{owl}
 and about two
orders of magnitude smaller at  $E = 3 \times 10^{19}$~eV.
 Because of the steep rise of the aperture with energy,
 OWL would see more events at  
$E \geq  1 \times 10^{20}$~eV than at  $E \geq  3 \times 10^{19}$~eV.
With a flux of 0.04 (km$^2$ sr s)$^{-1}$,
 OWL would collect 9200 events per year. 
The number of events within the cones of solid angle $\Omega$ we are 
studying is a fraction $f= \Omega/ (4\pi)=
(1- \cos\theta)/2$ of these 9200 events per year.
 In five years of operation the minimum asymmetry 
 OWL could detect is shown as a function of $\theta$ 
in Figs.~2 and 3 with the thicker solid (red) line labeled OWL.
In particular, OWL would detect 276 events per year in a cone
of aperture $\theta =20^\circ$. This corresponds to a detection of $N=1380$ 
events within the same cone in 5 years. Thus OWL could
detect stream asymmetries above 1.12, i.e.\ 
 could reach local overdensities due to the stream 
 $\chi \geq  0.12$. In terms of overdensity due to the stream, OWL's
 five-year reach is better than the reach of the combined Auger 
South and North in 10 years.

 In conclusion, EUSO would not be able to see the stream.
The Southern Auger site alone cannot study the 
stream asymmetry unless events below a 60$^\circ$ zenith
 angle can be analyzed effectively. In the hypothetical case in which
 zenith angles as large as
 85$^\circ$ could be reached,  in 10 years Auger South alone could
 test local overdensities in the stream of $\chi \geq 0.3$ (if
 $R_c=5$~kpc).  This is possible only if 
$\rho_{\rm h} \leq 0.3$~GeV/cm$^3$. For example, 
if $\rho_{\rm h} = 0.3$~GeV/cm$^3$, the
 minimum stream density testable by Auger South alone in 10 years
 would be $\rho_{\rm stream} =$ 0.09~GeV/cm$^3$, while if 
$\rho_{\rm h} = 0.2$~GeV/cm$^3$, it would be 
$\rho_{\rm stream} =$ 0.06~GeV/cm$^3$. 

The best chance
of detecting the Sgr leading tidal stream  in UHECR would come either
 with 5 years of observation with OWL or 10 years of observation with 
the combined Northern and Southern Auger sites. In fact, the combined
Northern and Southern Auger data, integrated for about 10 years, would
allow us to detect local stream overdensities of $\chi \geq 0.14$ (if
$R_c=5$~kpc). This means $\rho_{\rm stream} \geq$ 0.042~GeV/cm$^3$ if
$\rho_{\rm h} = 0.3$~GeV/cm$^3$, or 
$\rho_{\rm stream} >$ 0.028~GeV/cm$^3$ if $\rho_{\rm h} = 0.2$~GeV/cm$^3$.

The OWL data, integrated for about 5 years, would allow us to detect 
local stream overdensitites of $\chi \geq 0.12$  (if
$R_c=5$~kpc). This means $\rho_{\rm stream} \geq$ 0.036~GeV/cm$^3$ if
$\rho_{\rm h} = 0.3$~GeV/cm$^3$, or 
$\rho_{\rm stream} >$ 0.024~GeV/cm$^3$ if $\rho_{\rm h} = 0.2$~GeV/cm$^3$.

The numbers given here are only indicative and we do not attempt to provide
an estimate of their error.
Our conclusions depend on the assumed
UHECR fluxes which we took here to be at the level measured by AGASA.
 Clearly, the searches would become less sensitive for 
smaller fluxes. Our conclusions  also depend on the halo model, in that
the asymmetry due to the stream would decrease if the halo is flattened in
the direction of the stream.

\vspace{.3cm}

P.~G.\ would like to thank Benjamin Stokes for helpful
 information on the sky coverage of UHECR experiments. We thank the referee
for suggesting corrections and Dmitry Semikoz for usefull discussions.
The work of G~.G.\ was partially supported by DOE grant DE-FG03-91ER40662
and by NASA grant NAG5-13399.

\newpage

\begin{figure}
\epsfig{file=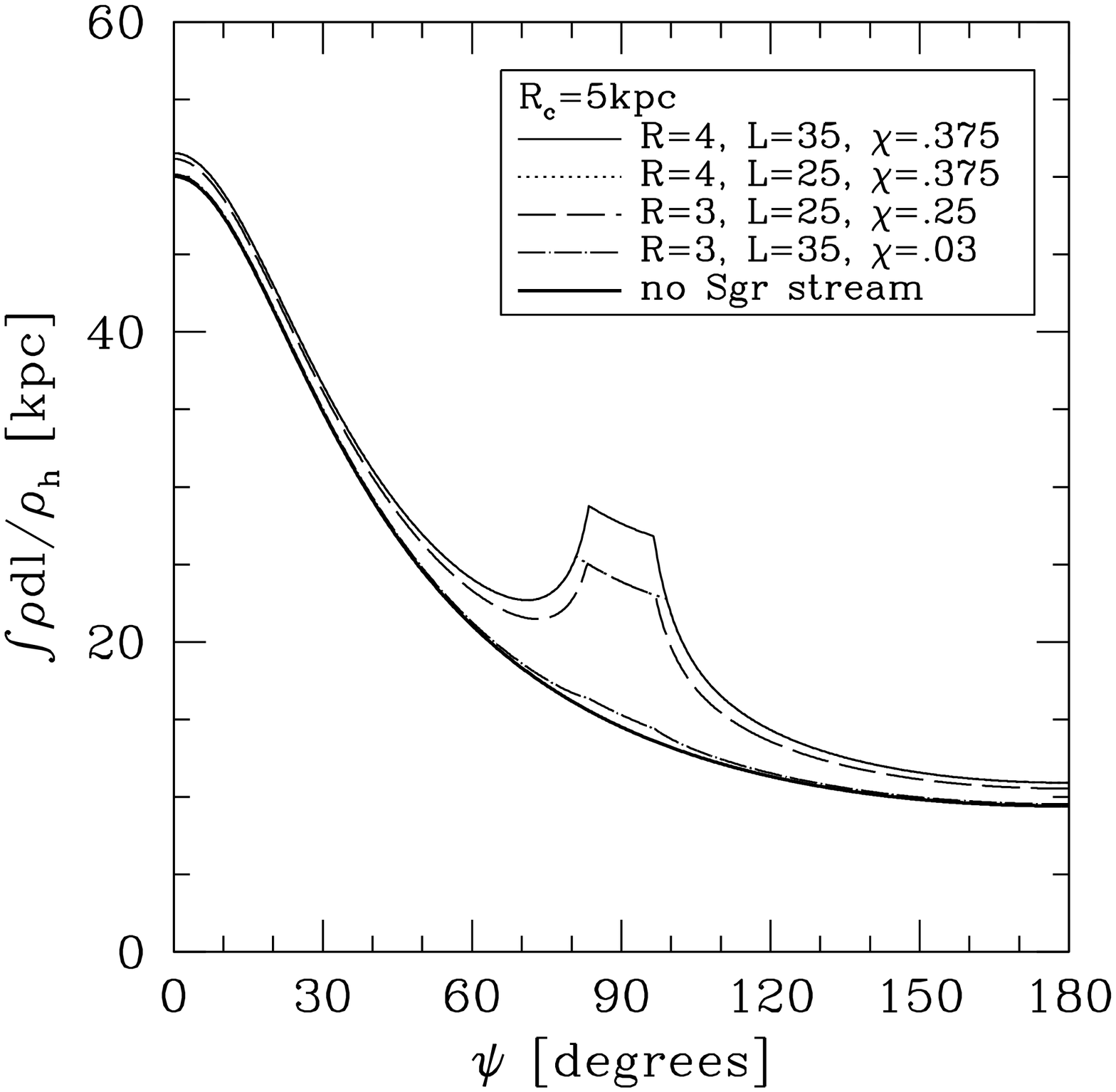,width=\textwidth}
\caption{Halo and stream column densities 
  for several stream models, for halo core radius $R_c=5$~kpc, as a
  function of the angle $\psi$ measured from the Galactic Center in
  the direction of the North Galactic Pole (the Galactic Center is at
  $\psi=0$, the North Galactic Pole is at $\psi=90^\circ$, and the
  Galactic Anticenter is at $\psi=180^\circ$).  }
\end{figure}
\newpage

\begin{figure}
\epsfig{file=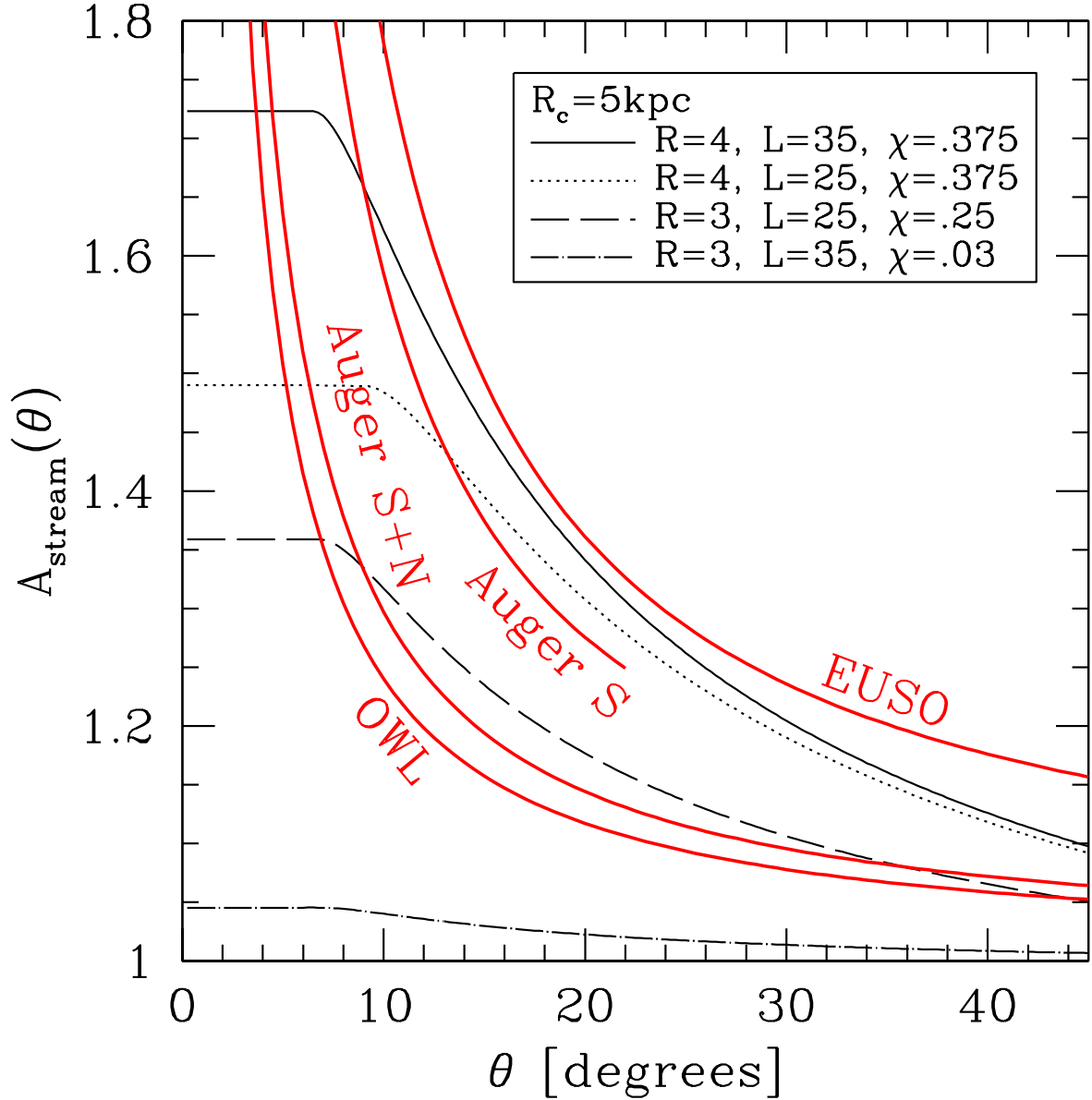,width=\textwidth}
\caption{ Asymmetry amplitudes $A_{\rm stream}(\theta)$ as a function of the
  cone aperture $\theta$, for $R_c$=5~kpc and several values of the
  overdensity $\chi$ due to the Sgr stream and the radius $R$ and
  length $L$ of the cylinder that models the Sgr stream. The thicker solid
  (red) superimposed lines indicate the reach of each of the
  observatories.  }
\end{figure}

\newpage

\begin{figure}
\epsfig{file=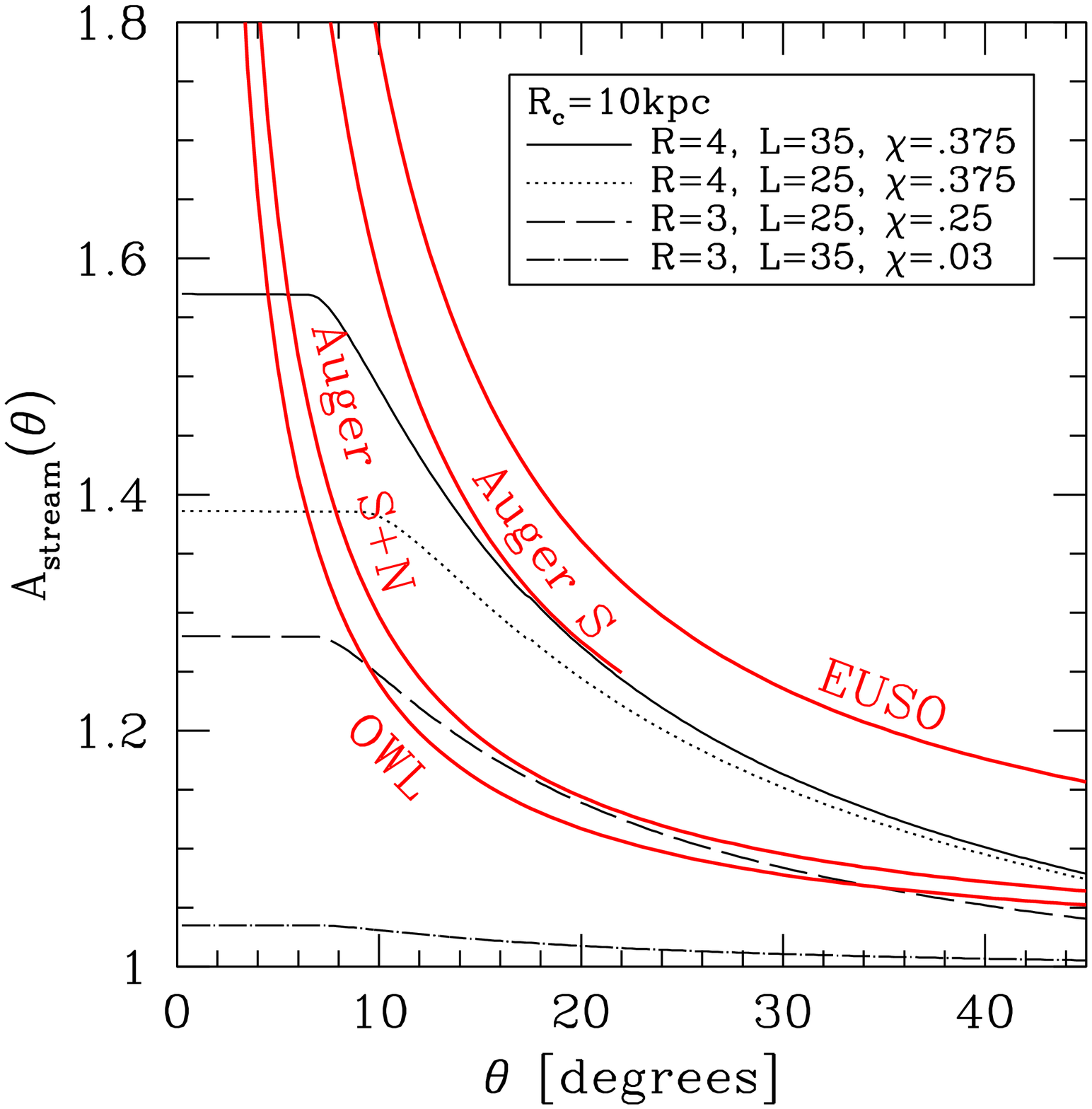,width=\textwidth}
\caption{ Same as Fig.~2 but for $R_c$=10~kpc.}

\end{figure}

\newpage

\begin{figure}
\epsfig{file=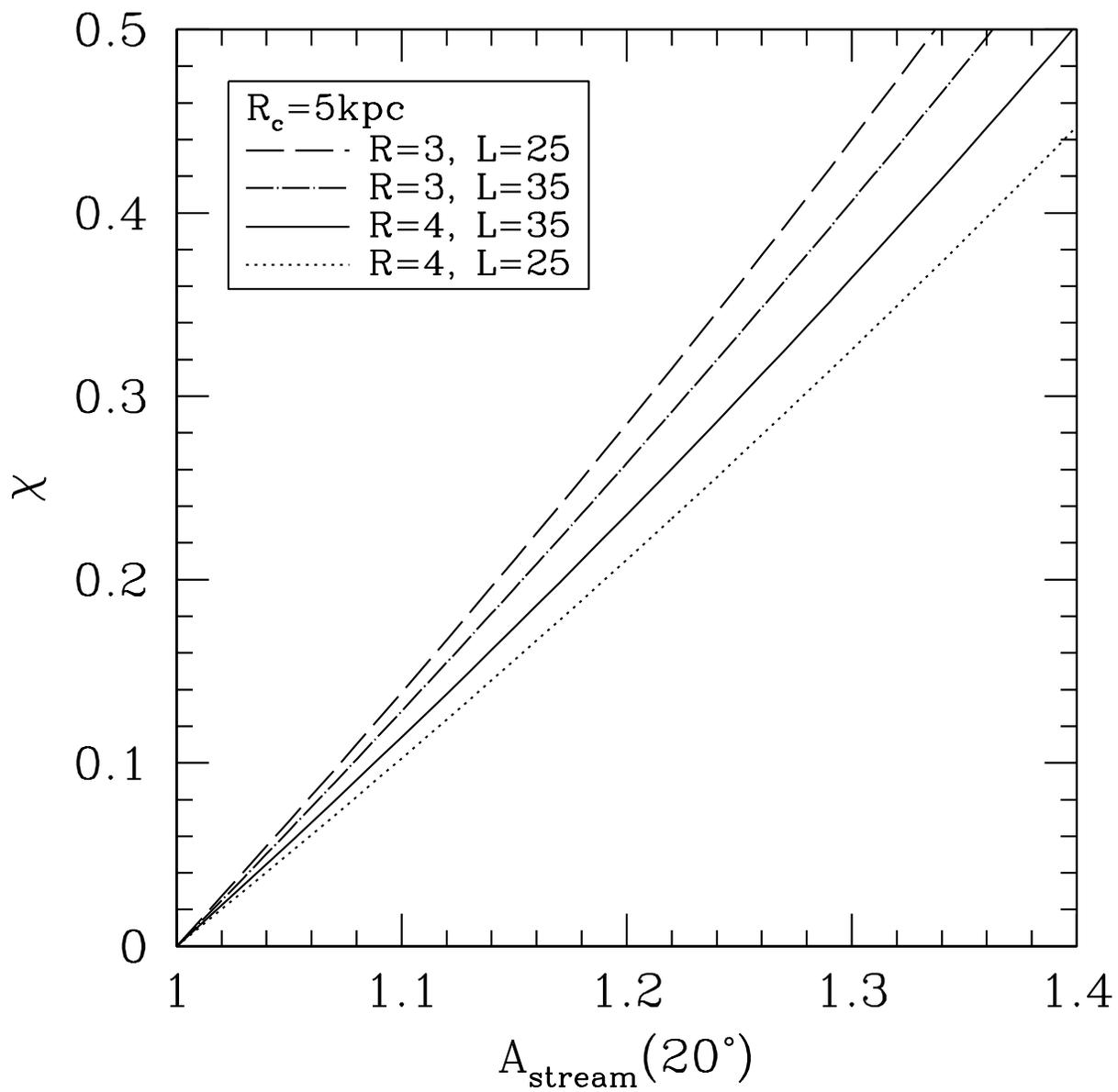,width=\textwidth}
\caption{Local overdensity due to the stream $\chi$  as a function of
the asymmetry amplitude for a cone of 20$^\circ$ aperture,
$A_{\rm stream}(20^\circ)$, for $R_c$=5 kpc.
}
\end{figure}

\newpage

\begin{figure}
\epsfig{file=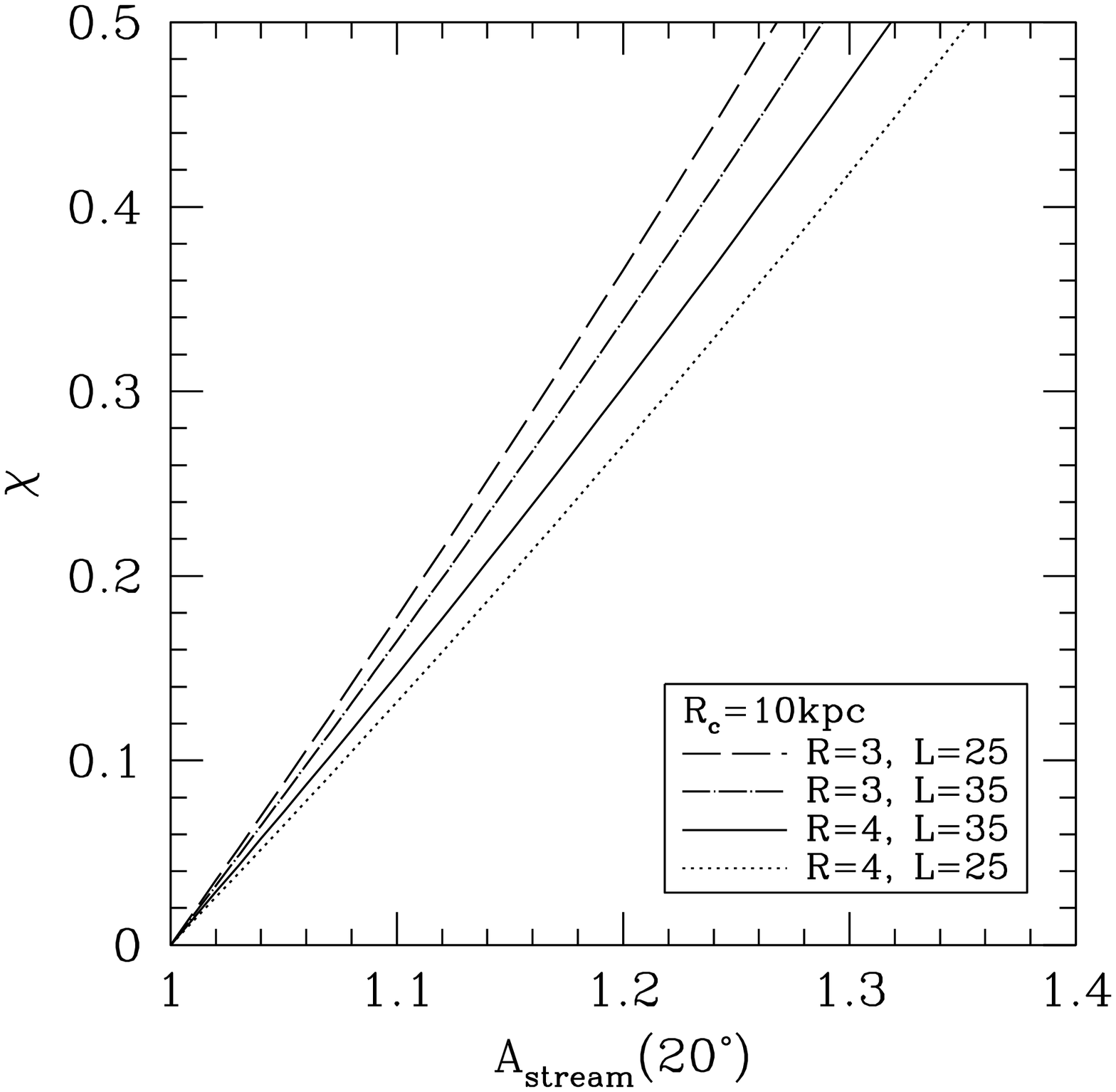,width=\textwidth}
\caption{
Same as Fig.~4 but for $R_c$=10 kpc.
}
\end{figure}

\end{document}